\documentclass[12pt]{iopart}

\usepackage{iopams}
\expandafter\let\csname equation*\endcsname\relax
\expandafter\let\csname endequation*\endcsname\relax 
\usepackage{graphicx}
\usepackage[latin1]{inputenc}
\usepackage{epsfig}
\usepackage{amsmath}
\usepackage{amsfonts}
\usepackage{float}
\usepackage{amssymb}
\usepackage{color}
\usepackage[colorlinks,citecolor=blue,urlcolor=blue,linkcolor=blue]{hyperref}

\begin{document}
	
\title{Explaining  $B$ decays anomalies in  SUSY models}

\author{Dris Boubaa$^{1,2}$, Shaaban Khalil$^{3}$ and Stefano Moretti$^{4}$}

\address{$^1$Department of Physics, Faculty of Exact Sciences and Computing,
	 Hassiba Benbouali University of Chlef, B.P 78C , Ouled Fares Chlef 02180, Algeria}
\address{$^2$Laboratoire de Physique des Particules et Physique Statistique, Ecole Normale Sup\'erieure-Kouba, 
           B.P. 92, 16050, Vieux-Kouba, Algiers, Algeria}
\address{$^3$Center for Fundamental Physics, Zewail City of
	Science and Technology, Sheikh Zayed,12588, Giza, Egypt}
\address{$^4$School of Physics and Astronomy, University of
	Southampton, Highfield, Southampton SO17 1BJ, UK}
\ead{d.boubaa@univ-chlef.dz, 
	skhalil@zewailcity.edu.eg, 
	s.moretti@soton.ac.uk. }
\vspace{10pt}

\begin{abstract}
Recent measurements of certain $B$ decays indicate deviations from Standard Model (SM) predictions.
We show that Supersymmetric effects can increase the Branching Ratios (BRs) of both $B \to D\tau\nu_\tau$ and $B \to D^*\tau\nu_\tau$
with respect to the SM rates, thereby approaching their newest experimentally measured values.
\end{abstract}
\section{Introduction}
\begin{figure}[t]
	\centering
	\includegraphics[height=7cm,width=10.cm]{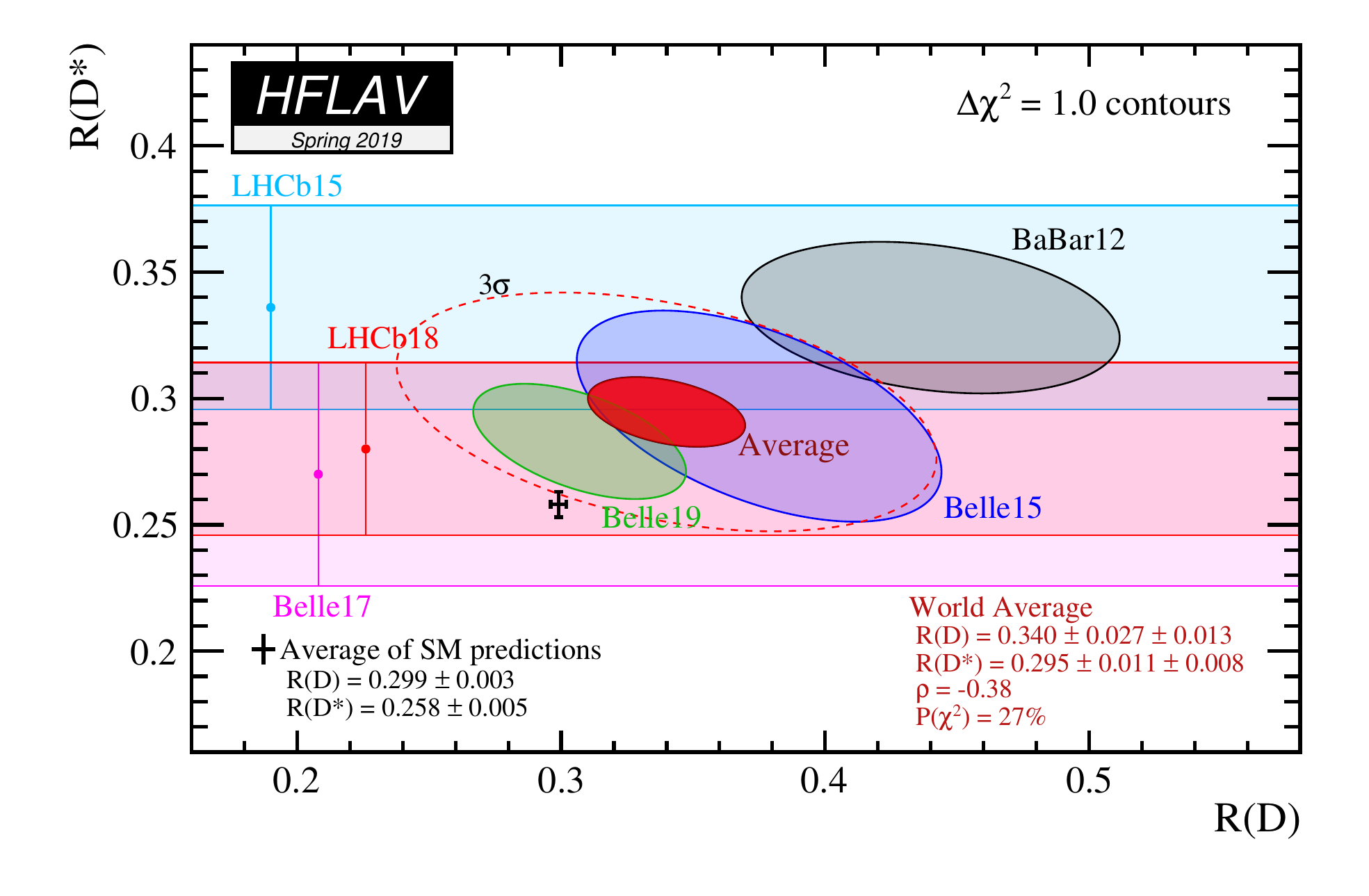}
	\vspace{-0.5cm}
	\caption{ Measurements of ${R}(D)$ and ${R}(D^*)$ reported in the last few years. The red ellipse shows the new world average while the dashed ellipse corresponds to a $3\sigma$ contour, i.e., 99.73\% Confidence Level (CL). The SM predictions are represented by  black bars. This {figure is taken from Ref.\ \cite{Amhis:2019ckw}.}}
	\vspace{-0.2cm}
	\label{FigAV}
\end{figure}

Semileptonic decays $B\to D^{(\ast)}\tau\bar\nu_\tau$ have been widely studied in the last few years which provide a good opportunity for testing the SM and searching for possible New Physics (NP) Beyond the SM (BSM). In fact, there are continuous efforts being undertaken at $B$ factories, so that the BaBar, Belle and LHCb collaborations continue to update their measurements with ever better precision. The ratios of semileptonic $B$-decay rates,
\begin{equation}
{R}(D)=\frac{{\rm BR}(\bar{B}\rightarrow D\tau\bar{\nu}_{\tau})}{{\rm BR}(\bar{%
		B}\rightarrow Dl\bar{\nu}_{l})}, ~~ {R}(D^{\ast})=\frac{{\rm BR}(%
	\bar{B}\rightarrow D^{\ast}\tau\bar{\nu}_{\tau})}{{\rm BR}(\bar{B}\rightarrow
	D^{\ast}l\bar{\nu}_{l})},~~(l=e,\mu),
\end{equation}
 have been measured by the three groups between 2012 and 2019. All measurements are shown in Fig.~\ref{FigAV}. Combining the experimental data for $ R(D)$ and $ R(D^*)$ using data from BaBaR \cite{Lees:2012xj,Lees:2013uzd}, Belle \cite{Huschle:2015rga,Sato:2016svk,Hirose:2016wfn,Hirose:2017dxl,Abdesselam:2019dgh} and LHCb \cite{Aaij:2015yra,Aaij:2017uff,Aaij:2017deq}, the Heavy Flavor Averaging  (HFLAV) Group determined the world averages for 2019 as \cite{Amhis:2019ckw} 
\begin{align}
{R}(D)&={0.340\pm0.027 \pm  0.013}, \label{average219}\\
{R}(D^*)&={0.295\pm0.011  \pm 0.008 },
\end{align}
which deviate by $1.4\sigma$ for ${R}(D)$ and $2.5\sigma$ for ${R}(D^*)$ from the SM expectations that are given by \cite{Amhis:2019ckw}
\begin{align}
{R}_{\rm SM}(D)&=0.299\pm0.003, \\
{R}_{\rm SM}(D^*) &=0.258\pm0.005,
\label{SM_pre} 
\end{align}
see Fig.~1.

 In this paper, based on \cite{Boubaa:2016mgn}, we argue that the recent experimental measurements of the so-called flavour anomalies $B\to D\tau\bar\nu_\tau$ and  $B\to D^{\ast}\tau\bar\nu_\tau$ can be explained by BSM physics. Specifically, we discuss that SUSY contributions, as described in
 the Minimal Supersymmetric Standard Model  (MSSM) with non-universal
 soft SUSY-breaking terms, might help to explain the discrepancy between the
 experimental results for $R(D)$  as well as $R(D^*)$ and the corresponding SM
 expectations. 
 
\section{${R(D)}$ and ${R(D^\ast)}$ in the MSSM}
The effective Hamiltonian for $b\rightarrow cl\bar{\nu_{l}}$ is
\begin{eqnarray}
{\mathcal{H}}_{\mathrm{eff}} &=& \frac{4G_{F}V_{cb}}{\sqrt{2}}\Big{[}%
(1+g_{VL})[\bar{c}\gamma_{\mu}P_{L}b][\bar{l}\gamma_{\mu}P_{L}\nu_{l}]
+ g_{VR} [\bar{c}\gamma_{\mu}P_{R}b][\bar{l}\gamma_{\mu}P_{L}%
\nu_{l}]\nonumber\\
& +&g_{SL} [\bar{c}P_{L}b][\bar{l}P_{L}\nu_{l}]  
+ g_{SR} [\bar{c}P_{R}b][\bar{l}P_{L}\nu_{l}]+g_{T} [\bar{c}\sigma^{\mu\nu_\tau
}P_{L}b][\bar{l}\sigma_{\mu\nu}P_{L}\nu_{l}]\Big{]},~~~~
\end{eqnarray}
where $G_{F}$ is the Fermi coupling constant,  $V_{cb}$ is the Cabibbo-Koboyashi-Maskawa (CKM) matrix element between charm and bottom quarks while $P_{L/R}=(1-/+\gamma_{5})/2$ are the chirality projection operators. 
Furthermore, $g_i$ is defined in terms of the Wilson coefficients  (see \cite{Bhattacharya:2015ida} for prospects of extracting 
these using optimal observables) $C_i$ as $%
g_{i}=C_{i}^{\mathrm{\rm SUSY}}/C^{\mathrm{SM}}$, with $i\equiv VL, VR, SL, SR, T$ and $C^{\mathrm{SM}}=\frac{4G_{F}V_{cb}}{\sqrt{2}}$.  
The amplitudes of possible NP contributions to $\bar{B}\rightarrow D^{(\ast)}l\bar{\nu}_{l}$, ${\cal M}\equiv \langle D^{(\ast)}l\bar{\nu}_{l} \vert  {\mathcal{H}}_{\mathrm{eff}} \vert \bar{B} \rangle$, can be written in the form \cite{Hagiwara:1989cu,Hagiwara:1989gza} 
\begin{align}
\mathcal{M}_{S(L,R)}^{\lambda_{D^{(\ast)}},\lambda_{l}}  &  \mathcal{=}%
\mathcal{-}\frac{G_{F}}{\sqrt{2}}V_{cb}g_{S(L,R)}H_{S(L,R)}^{\lambda_{D^{(\ast)}}%
}L^{\lambda_{l}},\\
\mathcal{M}_{V(L,R)}^{\lambda_{D^{(\ast)}},\lambda_{l}}  &  \mathcal{=}\frac{G_{F}%
}{\sqrt{2}}V_{cb}g_{V(L,R)}\sum_{\lambda}\eta_{\lambda}H_{V(L,R),\lambda
}^{\lambda_{D^{(\ast)}}}L_{\lambda}^{\lambda_{l}},\label{MVL}\\
\mathcal{M}_{T}^{\lambda_{D^{(\ast)}},\lambda_{l}}  &  \mathcal{=}\mathcal{-}%
\frac{G_{F}}{\sqrt{2}}V_{cb}g_{T}\sum_{\lambda,\lambda^{\prime}}\eta
_{\lambda^{\prime}}\eta_{\lambda}H_{\lambda\lambda^{\prime}}^{\lambda_{D^{(\ast)}}%
}L_{\lambda\lambda^{\prime}}^{\lambda_{l}}.
\end{align}
{The SM amplitude  is given by} 
{
	\begin{equation}
	\mathcal{M}_{\rm SM}^{\lambda_{D^{(\ast)}},\lambda_{l}}\mathcal{=}\frac{G_{F}}{\sqrt{2}%
	}V_{cb}\sum_{\lambda}\eta_{\lambda}H_{VL,\lambda}^{\lambda_{D^{(\ast)}}}L_{\lambda
}^{\lambda_{l}},
\end{equation}
where $\lambda_{l}$ is the helicity of the lepton $l$ and $\lambda,\lambda^{\prime}=\pm,0$ or $s$ are the helicity of virtual vector bosons. The $D^{(\ast)}$-meson is taken to be either a spin-0
$D$-meson, with $\lambda_{D}=s$, or a spin-1 $D^{\ast}$-meson, with
$\lambda_{D^{(\ast)}}=\pm,0$. The summation is over the virtual vector boson helicities
with the metric $\eta_{\pm}\ =\eta_{0}=-\eta_{s}=1$,$\ \ H$'s and $L$'s are
the hadronic and leptonic amplitudes which are defined in Refs.~\cite{Datta:2012qk,Duraisamy:2013kcw,Fajfer:2012vx,Crivellin:2012ye,Crivellin:2013wna}. Furthermore, one can also define the differential rate for the process $\bar{B}\rightarrow D^{(\ast)}l\bar{\nu}_{l}$ as
\begin{equation}
\frac{d\Gamma}{dq^{2}d\cos\theta_{l}}=\frac{\sqrt{Q_{+}Q_{-}}v_l}{256\pi^{3}%
	m_{B}^{3}}|\mathcal{M}(\bar{B}\rightarrow D^{(\ast)}l\nu_{l})|^{2},
\end{equation}
where $v_l=1-\frac{m_{l}^{2}}{q^{2}}$, $q^{2}$ varies in the range $m_{l}^{2}\leq q^{2}$
$\leq(m_{B}-m_{D^{(\ast)}})^{2}$, $Q_{\pm}=(m_{B}\pm m_{D^{(\ast)}})^{2}-q^{2}$ with $q^{\mu}=p_{B}^{\mu}-p_{D^{(\ast)}}^{\mu}=p_{l}^{\mu}+p_{\nu_{l}}^{\mu}$ and $-1\leq\cos\theta_{l}\leq1$.  Therefore, the full amplitude takes the form 
\begin{equation}
\mathcal{M}=\mathcal{M}_{\rm SM}^{\lambda_{D^{(\ast)}}%
	,\lambda_{l}}+\mathcal{M}_{S(L,R)}^{\lambda_{D^{(\ast)}},\lambda_{l}}%
+\mathcal{M}_{V(L,R)}^{\lambda_{D^{(\ast)}},\lambda_{l}}+\mathcal{M}_{T}%
^{\lambda_{D^{(\ast)}},\lambda_{l}}.
\end{equation}
Eventually, one can define both obsevables ${R}(D)$ and
${R}(D^*)$ as follows
\begin{equation}
{R}(D)=\frac{\Gamma(\bar{B}\rightarrow D\tau\nu_{\tau})}{\Gamma(\bar{B}\rightarrow Dl\nu_{l})},~~~{R}(D^{\ast})=\frac{\Gamma(\bar{B}\rightarrow D^{\ast}\tau\nu_{\tau})}{\Gamma(\bar{B}\rightarrow D^{\ast}l\nu_{l})}.
\end{equation}
Using the explicit formulae of the hadronic and leptonic amplitudes in Refs.~\cite{Hagiwara:1989cu,Hagiwara:1989gza,Datta:2012qk,Duraisamy:2013kcw,Tanaka:2012nw,Sakaki:2013bfa,Sakaki:2014sea} (when the $l$ contribution is assumed to be described by the SM) and upon fixing the SM parameters and
the form factors involved in the definition of the matrix elements
to their central values as in Ref.~\cite{Lees:2013uzd}, we can cast
the explicit dependence of  ${{R}}(D)$ and
${{R}}(D^*)$ upon the Wilson coefficients in the MSSM as follows \cite{Boubaa:2016mgn}:
	\begin{align}
	R(D)&  ={R}(D)^{\mathrm{SM}}\Big[0.981|g_{SR}+g_{SL}|^{2}%
	+|1+g_{VL} +g_{VR}|^{2} +0.811 |g_{T}|^{2}\nonumber\\ &+1.465 \operatorname{Re}[(1+g_{VL}+g_{VR})
	\times(g_{SR}+g_{SL})^{\ast}]
	+1.074\operatorname{Re}[(1+g_{VL}+g_{VR})g_{T}^{\ast}]\Big],\\
	R({D^{\ast}}) &  ={R}(D^{\ast})^{\mathrm{SM}}\Big[0.025 |g_{SR}%
	-g_{SL}|^{2}+|1+g_{VL}|^{2}+|g_{VR}|^{2}
	+16.739|g_{T}|^{2}\nonumber\\
	&+0.094\operatorname{Re}[(1+g_{VL}+g_{VR})
	\times(g_{SR}-g_{SL})^{\ast}]
	+6.513\operatorname{Re}[g_{VR}g_{T}^{\tau\ast}]\nonumber\\
	&-4.457\operatorname{Re}[(1+g_{VL})g_{T}^{\ast}] -1.748\operatorname{Re}%
	[(1+g_{VL})g_{VR}^{\ast}]\Big].
	\end{align}
	\begin{figure}[!ht]
		\centering
		\includegraphics[height=5cm,width=7.cm]{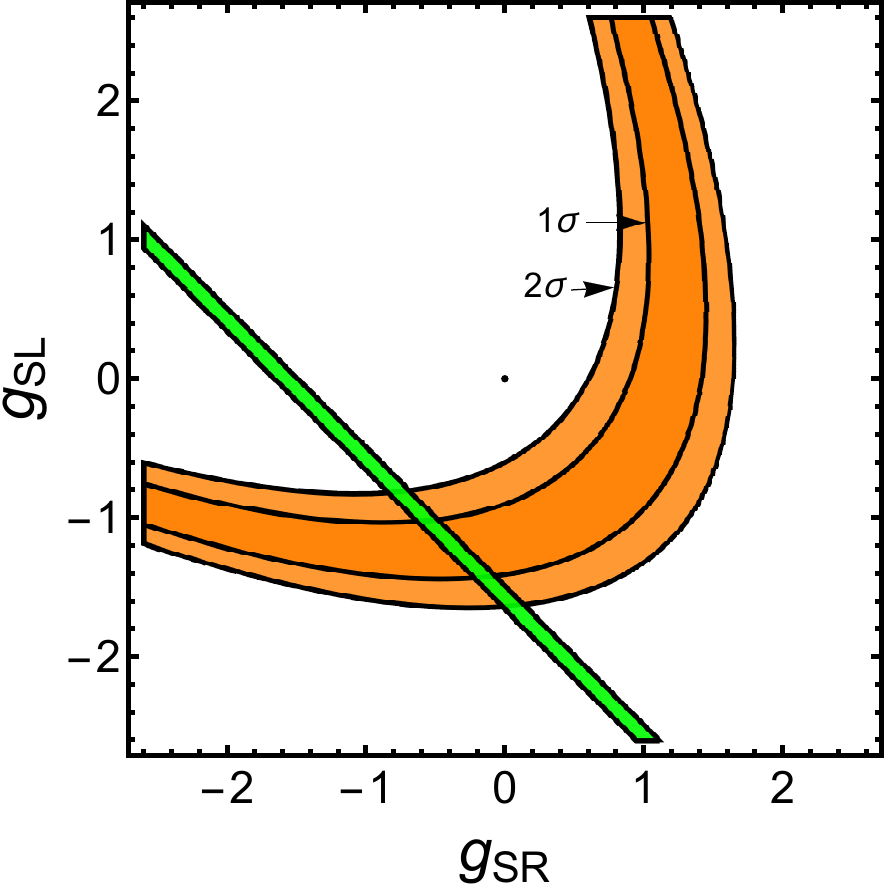}~~~ \includegraphics[height=5 cm,width=7.cm]{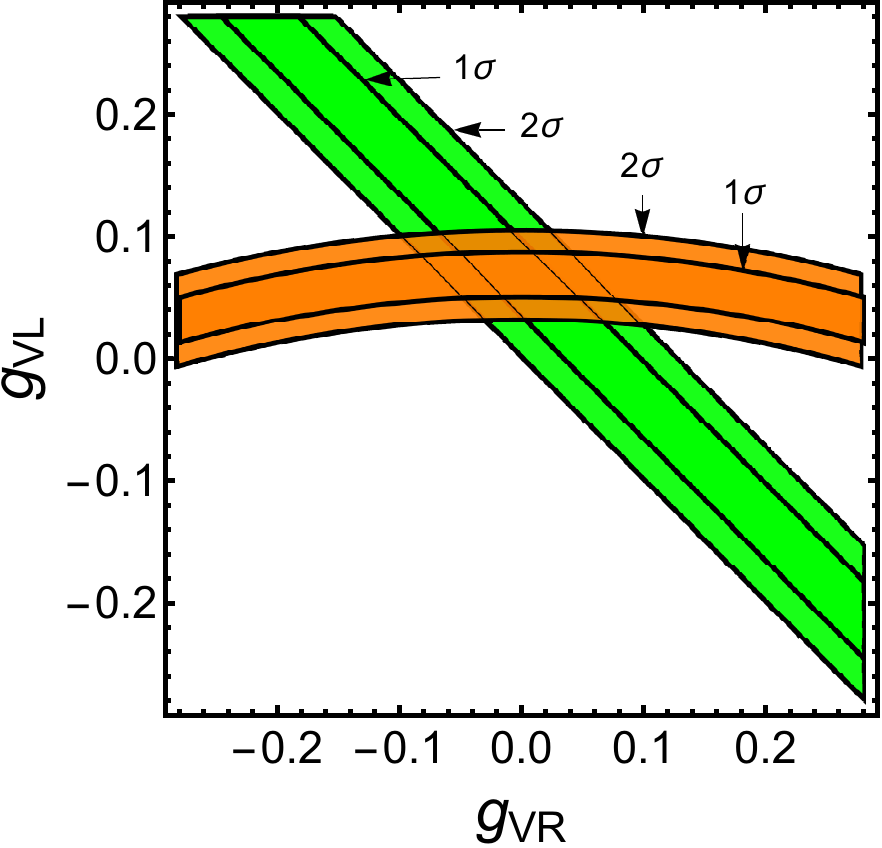}	
		\vspace{-0.1cm}
		\caption{The allowed regions in the $(g_{SL}, g_{SR})$  (left) and  $(g_{VL}, g_{VR})$  (right) planes by the $1\protect\sigma$ and $2\protect\sigma$ experimental results on ${R}(D)$ (green) and ${R}(D^*)$ (orange) of the  2019 averages.}
		\label{fig1}
	\end{figure}

Thus, in case of a dominant scalar contribution (and negligible vector and
tensor ones), it is clear that ${R}(D^*)$ cannot be significantly larger than
the SM expectation, due to the smallness of the  coefficient of  this contribution, unless $\vert g_{SR} - g_{SL}\vert $ is much larger than 1 ({i.e.}, $C_S^{\mathrm{\rm SUSY}} > C^{\mathrm{SM}}$), which is not possible.
This conclusion is confirmed in Fig.~\ref{fig1}, where we display the regions in the ($g_{SL}, g_{SR}$) plane that can accommodate the experimental
results of ${R}(D)$ and ${R}({D^*})$ within $1 \sigma$ and $2 \sigma$  CL for, e.g., Belle, the experiment with predictions closer to the SM.
From this figure, it is clear that the scalar contribution alone cannot account for both ${R}(D)$ and ${R}({D^*})$
simultaneously. In order to get ${R}(D)$ and ${R}({D^*})$ within $2\sigma$ of the  
aforementioned average results from the various experiments, $(g_{SL}, g_{SR})$ should lie between
$(-0.75, -0.69)$ and $(-0.04, -1.65)$, respectively. In these conditions, either $g_{SL}$ or $g_{SR}$ is larger than 1, which is not possible.

In case of a dominant vector contribution, as shown from the allowed regions of  ($g_{VL}, g_{VR}$) in Fig.~\ref{fig1}, one gets ${R}(D)$ and ${R}(D^*) $ inside the $2 \sigma$ region of the averages if $(g_{VL}, g_{VR})$ varies between $(0.03,-0.03)$ and $(0.1,0.02)$, respectively.
Furthermore, it is remarkable that, unlike the scalar contribution, a small vector contribution, $g_{VL}$ $\sim$ $ {\cal O}(0.1)$ and $g_{VR} \sim {\cal O}(0.01)$, can induce significant enhancement for both ${R}(D) $ and ${R}(D^*) $: e.g.,  ${R}(D) \sim 0.336$ and ${R}(D^*) \sim 0.277$ if $g_{VL} \sim 0.05$ and $g_{VR} \sim 0$, which, as we will see, are quite plausible values in the MSSM. Finally, the tensor contribution, which is typically quite small, may affect only ${R}(D^*)$.

The SUSY contributions to $g_{VL}$ are generated from the penguin corrections to the vertex $W^\pm l \nu_l $ $(l=e,\mu,\tau)$ through the exchange of charginos and neutralinos alongside sleptons and sneutrinos, respectively, as displayed in Fig.~\ref{fig2}. Let us now try to decode our results, by concentrating on the Wilson coefficient $C_{VL}$, which sees contributions induced by the penguin topologies in Fig. \ref{fig2}.
Firstly, we can confirm that the graph with neutral Higgs bosons is small while the other two are roughly comparable. Thus, the emerging $C_{VL}^{\mathrm{SUSY}}$ term is essentially 

%
\begin{figure}[!t]
	\centering \epsfig{file=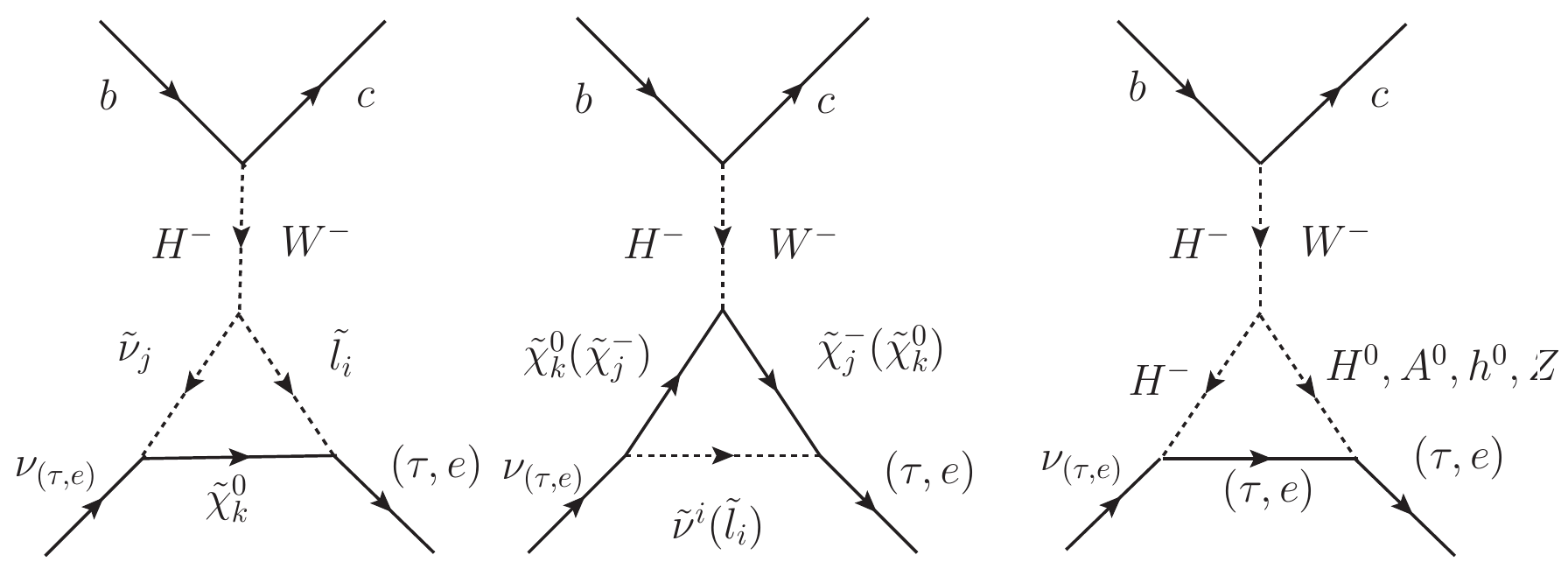,height=3.5cm,width=10cm,angle=0}
	\caption{Triangle diagrams (penguins) contributing to, {e.g.}, $b
		\to c \protect(\tau,e) \protect\nu_{(\tau,e)}$ affecting the leptonic vertex.
	}
	\label{fig2}
\end{figure}


\begin{equation}\label{CVL1}
C_{VL}^{\mathrm{SUSY}}=C_{VL}^{\tilde{%
		\tau}}+C_{VL}^{\tilde{\nu}} +C_{VL}^{(A^{0},H^{0},h^0)},
\end{equation}%
where 
\begin{align}\label{CVL}
C_{VL}^{\tilde{\tau}} &=
\frac{\Gamma_{\tilde{\chi}_{j}^{-}\nu_{l_{I}}\tilde{\tau}^{\ast}_{i}}^{L}
	\Gamma_{\bar{\l}_{I}\tilde{\chi}_{k}^{0}\tilde{\tau}_{i}}^{R}\Gamma_{\bar{c
		}bW^{-}}^{L}}{
	16\pi^{2}M_{W^\pm}^{2}}\Big{[}\Gamma_{\tilde{\chi}_{j}^{+}\chi_{k}^{0}W^{-}}^{R}m_{\tilde{\chi}_{j}^{-}}m_{\tilde{\chi}_{k}^{0}}  C_{0}(m_{\tilde{\chi}_{k}^{0}}^{2},m_{\tilde{\chi}_{j}^{-}}^{2},m_{\tilde{\tau}_{i}}^{2})\nonumber\\
&-\Gamma_{\tilde{\chi}_{j}^{+}\tilde{\chi}_{k}^{0}W^{-}}^{L}(B_{0}(m_{\tilde{\chi}_{j}^{-}}
^{2},m_{\tilde{\chi}_{k}^{0}}^{2}) 
-2C_{00}(m_{\tilde{\chi}_{k}^{0}}^{2},m_{\tilde{\chi}_{j}^{-}}^{2},m_{\tilde{\tau}_{i}}^{2}) + m_{\tilde{\tau}_{i}}^{2}C_{0}(m_{\tilde{\chi}_{k}^{0}}^{2},m_{\tilde{\chi}_{j}^{-}}^{2},m_{\tilde{\tau}_{i}}^{2}))\Big{]},
\\
C_{VL}^{\tilde{\nu}} &=
\frac{\Gamma_{\nu_{l_{I}}\tilde{\chi}_{k}^{0}\tilde{\nu}_{i}^{\ast}}^{L}\Gamma_{\tilde{\chi}_{j}^{-}\bar{l_{I}}\tilde{\nu}_{i}}^{R}	
	\Gamma_{\bar{c}bW^{-}}^{L}}{
	16\pi^{2}M_{W^\pm}^{2}}\Big{[}-\Gamma_{\tilde{\chi}_{j}^{+}\chi_{k}^{0}W^{-}}^{L}m_{\tilde{\chi}_{j}^{-}}m_{\tilde{\chi}_{k}^{0}}  C_{0}(m_{\tilde{\chi}_{j}^{-}}^{2},m_{\tilde{\chi}_{k}^{0}}^{2},m_{\tilde{\nu}_{i}}^{2})\nonumber\\
&+\Gamma_{\tilde{\chi}_{j}^{+}\tilde{\chi}_{k}^{0}W^{-}}^{R}(B_{0}(m_{\tilde{\chi}_{k}^{0}}^{2},m_{\tilde{\chi}_{j}^{-}}
^{2}) 
-2C_{00}(m_{\tilde{\chi}_{j}^{-}}^{2},m_{\tilde{\chi}_{k}^{0}}^{2},m_{\tilde{\nu}_{i}}^{2}) + m_{\tilde{\tau}_{i}}^{2}C_{0}(m_{\tilde{\chi}_{j}^{-}}^{2},m_{\tilde{\chi}_{k}^{0}}^{2},m_{\tilde{\nu}_{i}}^{2}))\Big{]},
\\
C_{VL}^{A^{0}}  &=
\frac{2\Gamma_{\bar{\l} \nu_{\l} H^{-}}^{L}
	\Gamma_{\bar{\l}\l A^{0}}^{R}
	\Gamma_{A^{0}H^{+}W^{-}}
	\Gamma_{\bar{c
		}bW^{-}}^{L}}{
	16\pi^{2}M_{W^\pm}^{2}} C_{00}(m_{\l}^{2},M_{H^{-}}^{2},m_{A^{0}}^{2}).
\end{align}
The Wilson coefficients $C_{VL}^{(H^0,h^0)}$ can be obtained from $C_{VL}^{A^0}$ by exchanging $A^{0}\leftrightarrow (H^{0},h^0)$.
The corresponding couplings are given by
\begin{align}
\Gamma_{\tilde{\chi}_{j}^{-}\nu_{l_{I}}\widetilde{\tau}_{i}^{\ast}}^{L}  &= g(-Z_{L}^{iI\ast}
Z_{-}^{j1\ast}+\frac{m_{\l_{I}}}{\sqrt{2}M_{W^\pm}\cos\beta}Z_{L}^{i(I+3)\ast}Z_{-}^{j2
}), \label{chargino_coup}\\
\Gamma_{\bar{\l}_{I}\tilde{\chi}_{k}^{0}\widetilde{\tau}_{i}}^{R}  & =\frac{g}{\sqrt{2}%
}(Z_{L}^{iI\ast}(\tan\theta_{W}Z_{N}^{k1\ast}+Z_{N}^{k2\ast}) -\frac{m_{l_{I}}}{M_{W^\pm}\cos\beta}Z_{L}^{i(I+3)\ast}Z_{N}^{j3\ast}),\\
\Gamma_{\tilde{\chi}_{k}^{0}\nu_{l_{I}}\tilde{\nu}_{i}}^{L}  & =\frac{g}{\sqrt{2}}Z_{\nu}^{iI\ast}(\tan\theta_{W}Z_{N}^{k1}-Z_{N}^{k2}),~~~~~~~\Gamma_{\overline{c}bW^{+}}^{L} =-\frac{g}{\sqrt{2}}V_{cb},\\
\Gamma_{\bar{\l}_{I}\chi_{j}^{-}\tilde{\nu}_{i}}^{R}  & =- gZ_{+}^{j1\ast}Z_{\nu}^{iI},~~~~
\Gamma_{\tilde{\chi}_{j}^{+}\tilde{\chi}_{k}^{0}W^{-}}^{L}   =-g(Z_{-}^{j1}Z_{N}%
^{k2\ast}+\frac{1}{\sqrt{2}}Z_{-}^{j2}Z_{N}^{k3\ast}),\\
\Gamma_{\tilde{\chi}_{j}^{+}\tilde{\chi}_{k}^{0}W^{-}}^{R}  & =-g(Z_{+}^{j1}Z_{N}%
^{k2\ast}-\frac{1}{\sqrt{2}}Z_{+}^{j2}Z_{N}^{k4\ast}), ~~~\Gamma_{A^{0}H^{+}W^{-}} = \frac{g}{2}\\
\Gamma_{\bar{\tau} \nu H^{-}}^{L} & = \frac{g m_{\tau}}{\sqrt{2}M_{W^\pm}\cos\beta}Z_{H^{-}}^{21},~~~\Gamma_{\bar{\tau}\tau A^{0}}^{R}  =- \frac{1}{\sqrt{2}}\frac{g m_{\tau}}{\sqrt{2}M_{W^\pm}\cos\beta}Z_{A}^{21}, \\
\Gamma_{\bar{\tau}\tau H^{0}}^{R} & = \frac{1}{\sqrt{2}}\frac{g m_{\tau}}{\sqrt{2}M_{W^\pm}\cos\beta}Z_{H}^{21},~~
\Gamma_{H^{0}H^{+}W^{-}}  = \frac{g}{2}(Z_{H}^{22}Z_{H^{-}}^{22}-Z_{H}^{21}Z_{H^{-}}^{21}),\label{neutralino_coup}
\end{align}
\begin{figure}[!t]
	\centering
	\includegraphics[height=5cm,width=8cm]{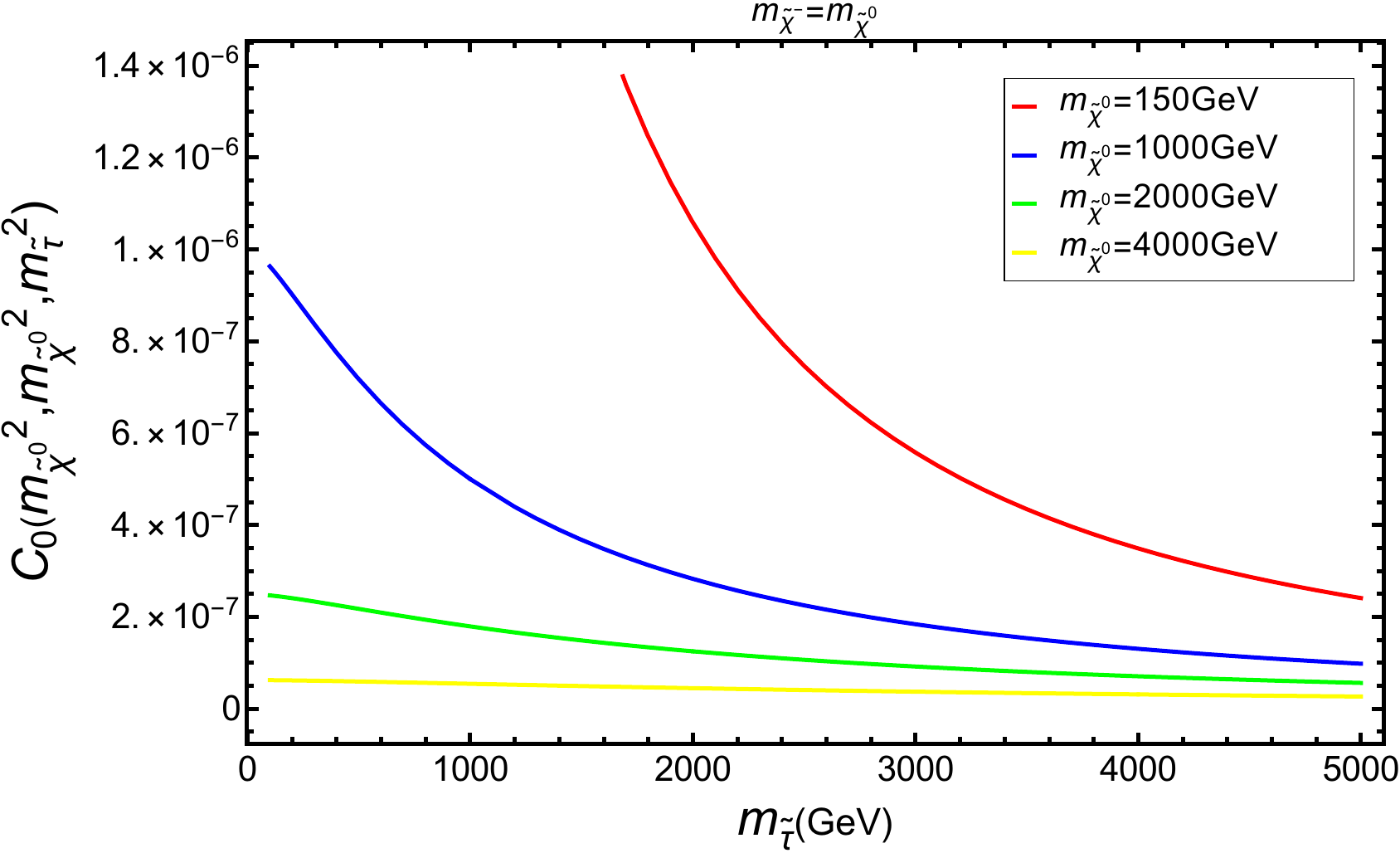}~~~
	\includegraphics[height=5cm,width=8cm]{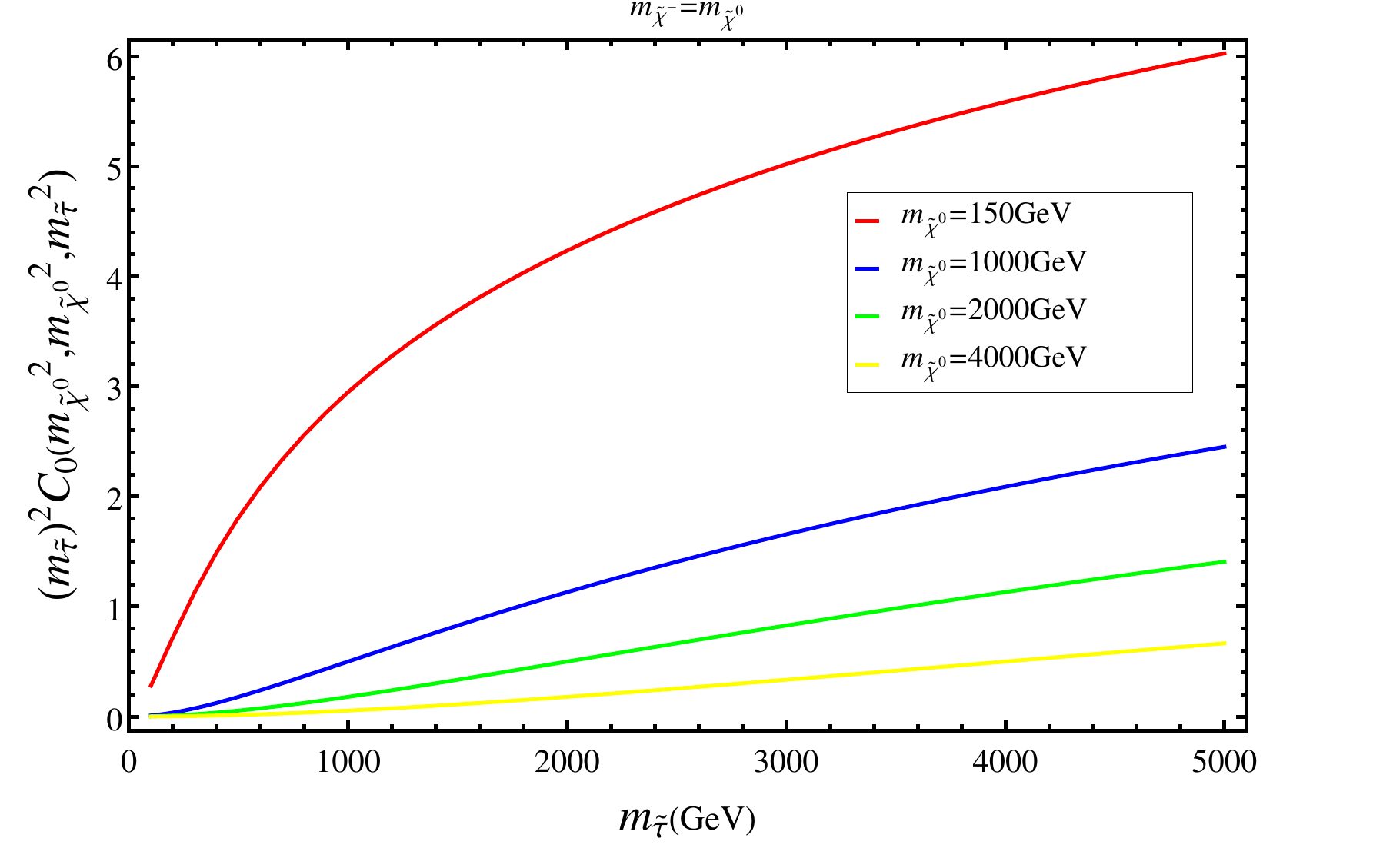}~~~
	\vspace{-0.3cm}
	\caption{Behaviour of the last term in Eq.~(\ref{CVL}), $m_{\tilde{\tau}}^2C_0(m_{\tilde{\chi}^{0}}^{2},m_{\tilde{\chi}^{-}}^{2},m_{\tilde{\tau}}^{2})$(right) and $C_0(m_{\tilde{\chi}^{0}}^{2},m_{\tilde{\chi}^{-}}^{2},m_{\tilde{\tau}}^{2})$(left), with $m_{\tilde{\tau}}$ for degenerate  chargino/neutralino masses.}
	\label{figCo}
\end{figure}
where $Z_L$, $Z_\nu$, $Z_\pm$, $Z_N$ and $Z_{(H,A,H^{-})}$ are the diagonalising matrices for
slepton, sneutrino, chargino, neutralino and Higgs masses, respectively.
In addition, the loop functions are given by \cite{Buras:2002vd}

\begin{align}
B_{0}(x,y)& =\eta _{\varepsilon }-1+\log \frac{x}{{\tilde\mu} ^{2}}-\frac{y\log
	\frac{y}{x}}{x-y}, \\
C_{0}(x,y,z)& =\frac{1}{y-z}\Big{(}\frac{y\log \frac{y}{x}}{y-x}+\frac{z\log
	\frac{z}{x}}{x-z}\Big{)}, \\
C_{00}(x,y,z)& =\frac{1}{4}\Big{(}\eta _{\varepsilon }-\log \frac{x}{\tilde{\mu} ^{2}
}\Big{)}+\frac{3}{8}
+\frac{1}{y-z}\Big{(}\frac{y^{2}\log \frac{y}{x}}{4(x-y)}-\frac{z^{2}\log\frac{z}{x}}{4(x-z)}\Big{)},
\end{align}
with $\eta _{\varepsilon }=\frac{2}{d-4 }+\log 4\pi \gamma_{E} $, which is subtracted in the modified 
Dimensional Regularisation/Reduction  
($\overline{\rm DR}$) scheme, and $\tilde \mu$  the renormalisation scale with the dimensions of mass. Our calculation  is based on  FlavorKit \cite{Porod:2014xia}, SARAH \cite{Staub:2013tta} and SPheno
\cite{Porod:2011nf}. Here, a few comments are in order. $(i)$ The loop function $C_{0}(m_{\tilde{\chi}_{k}^{0}}^{2},m_{\tilde{\chi}_{j}^{-}}^{2},m_{\tilde{\tau}_{i}}^{2}) \to 0$ if $m_{\tilde{\chi}_{k}^{0}}, m_{\tilde{\chi}_{j}^{-}}$ and $m_{\tilde{\tau}_{i}} \to \infty$, as expected in the SUSY decoupling limit. $(ii)$ If $m_{\tilde{\chi}_{k}^{0}}$ and $ m_{\tilde{\chi}_{j}^{-}}$ are of order ${\cal O}(100)$ GeV and $m_{\tilde{\tau}_{i}} $ is very heavy, { then $m_{\tilde{\tau}_{i}}^2 C_{0}(m_{\tilde{\chi}_{k}^{0}}^{2},m_{\tilde{\chi}_{j}^{-}}^{2},m_{\tilde{\tau}_{i}}^{2})$ does not vanish, as this is not a decoupling limit since a light fermionic SUSY spectrum is assumed}. Specifically, for
$m_{\tilde{\chi}_{k}^{0}} \simeq m_{\tilde{\chi}_{j}^{-}}$, the loop function takes the form
\begin{equation} 
C_0(m^2_{\tilde{\chi}_{i}^{0}},m^2_{\tilde{\chi}_{i}^{0}}, m^2_{\tilde{\tau}_{j}}) = \frac{1}{(m^2_{\tilde{\chi}_{i}^{0}} -m^2_{\tilde{\tau}_{j}})^2} \left[ m^2_{\tilde{\chi}_{i}^{0}} - m^2_{\tilde{\tau}_{j}} + m^2_{\tilde{\tau}_{j}} \log\left( \frac{m^2_{\tilde{\tau}_j}}{m^2_{\tilde{\chi}_{i}^{0}}}\right)\right].
\end{equation}
$(iii)$ From Eq.~(\ref{CVL}), one can see that, if $C_0(m^2_{\tilde{\chi}_{i}^{0}},m^2_{\tilde{\chi}_{i}^{0}}, m^2_{\tilde{\tau}_{j}}) \neq 0$, then the last term, proportional to $m^2_{\tilde{\tau}_{j}} C_0(m^2_{\tilde{\chi}_{i}^{0}},m^2_{\tilde{\chi}_{i}^{0}},$ $m^2_{\tilde{\tau}_{j}})$, gives the dominant effect to $C_{VL}^{\tilde{\tau}}$. These comments are explicitly displayed in  Fig. \ref{figCo}. Thus, the typical values of the couplings $\Gamma_{\tilde{\chi}_{j}^{-}\bar{l_{I}}\tilde{\nu}_{i}}^{R}$, 	$\Gamma_{\nu_{l_{I}}\tilde{\chi}_{k}^{0}\tilde{\nu}_{i}^{\ast}}^{L}$,
$\Gamma_{\bar{c}bW^{-}}^{L}$, $\Gamma_{\tilde{\chi}_{j}^{+}\tilde{\chi}_{k}^{0}W^{-}}^{L}$ and the loop function $C_0(m^2_{\tilde{\chi}_{i}^{0}},m^2_{\tilde{\chi}_{i}^{0}}, m^2_{\tilde{\tau}_{j}}) $ at $m_{\tilde{\chi}_{i}^{0}} \sim {\cal O}(100)$ GeV and
$ m_{\tilde{\tau}_{j}} \sim {\cal O}(1)$ TeV imply that $C_{VL}^{\tilde{\tau}} \sim
\frac{2\times10^{-3}}{16\pi^2M_{W^\pm}^2} m^2_{\tilde{\tau}_{j}} C_0(m^2_{\tilde{\chi}_{i}^{0}},m^2_{\tilde{\chi}_{i}^{0}}, m^2_{\tilde{\tau}_{j}})$ is of order $10^{-8}~{\rm GeV}^{-2}$. Therefore, $g_{VL} = C_{VL}^{\tilde{\tau}} /C^{\rm SM}$, where $C^{\rm SM}\sim 1.38\times 10^{-6} ~{\rm GeV}^{-2}$, can be of order $0.01$.

Finally, one should consider a possible constraint due to the direct measurement of the $W^\pm$ boson decay widths that leads to \cite{Olive:2016xmw}
\begin{equation}\label{eq:Wen}
\Gamma(W\rightarrow\tau\nu)/\Gamma(W\rightarrow e\nu)=1.043\pm 0.024.
\end{equation}
The SM prediction for this ratio is given by $\sim 0.999267$, which is consistent with the measured value.  {{Similarly, constraints can also be obtained from}} \cite{Olive:2016xmw}
\begin{equation}\label{eq:Wmun}
\Gamma(W\rightarrow\tau\nu)/\Gamma(W\rightarrow \mu\nu)=1.07\pm 0.026,
\end{equation}
{{with which the SM is also consistent. {{Another important experimental measurement connected with  lepton universality in $\tau$ decay that should be considered here
is of $\tau \to \nu_{\tau}l\nu_{l}$  with $l=e,\mu$, which is given by the relation \cite{Aubert:2009qj}
	\begin{equation}\label{LM}
	\left(\frac{g_{\mu}}{g_{e}}\right)^{2}_\tau=\frac{BR(\tau \to \mu\nu_{\tau}\nu_{\mu})}{BR(\tau \to e\nu_{\tau}\nu_{e})}\frac{f(m_{e}^{2}/m_{\tau}^{2})}{f(m_{\mu}^{2}/m_{\tau}^{2})}.
\end{equation}
In the SM, the universal gauge interaction implies that
\begin{equation}\label{LU}
\frac{\Gamma(\tau \to \mu\nu_{\tau}\nu_{\mu})}{\Gamma(\tau \to e\nu_{\tau}\nu_{e})}=\frac{f(m_{\mu}^{2}/m_{\tau}^{2})}{f(m_{e}^{2}/m_{\tau}^{2})}=0.9726,
\end{equation}
where $f(x)=1-8x+8x^{3}-x^{4}-12x^{2}\log(x)$. The current experimental result for this ratio is $0.979\pm 0.004$ \cite{Olive:2016xmw}, which gives $\left(\frac{g_{\mu}}{g_{e}}\right)_\tau=1.0032\pm0.002$. With SUSY contributions, Eq.~(\ref{LU}) can be written as
\begin{equation}\label{LN}
\frac{\Gamma(\tau \to \mu\nu_{\tau}\nu_{\mu})}{\Gamma(\tau \to e\nu_{\tau}\nu_{e})}=0.9726\frac{|1+g_{VL}^{\mu}|^{2}}{|1+g_{VL}^{e}|^{2}},
\end{equation}
where $g_{VL}^{l}=C^{\rm SUSY}(\tau \to \nu_{\tau}l\nu_{l})/C^{\rm SM}(\tau \to \nu_{\tau}l\nu_{l})$ with $C^{\rm SM}(\tau \to \nu_{\tau}l\nu_{l})=2\sqrt{2}G_{F}$. (As we will show, this imposes stringent constraints on  SUSY contributions to $g_{VL}^l$)}.
{{Furthermore, SUSY loop effects induce a correction to the Fermi coupling via a potential breaking of $\mu-e$ universality. In fact, using Eqs.~(\ref{LM}) and (\ref{LN}), {for $g_{VL}^l\ll 1$ one can find}
\begin{equation}\label{eq:GF}
\left(\frac{g_{\mu}}{g_{e}}\right)_\tau=\frac{|1+g_{VL}^\mu|}{|1+g_{VL}^e|}=|1+\Delta g_{VL}^{\mu,e}|,
\end{equation}
{where $\Delta g_{VL}^{\mu,e}=g_{VL}^\mu-g_{VL}^e$, so that the above experimental constraints impose that $0.0012\leq\Delta g_{VL}^{\mu,e}\leq 0.0052$.
In our work, we will enforce $g_{\mu}=g_e=g$, which  satisfies Eq.~(\ref{eq:GF}). }
	
{{Furthermore, the oblique Electro-Weak
(EW) parameters $S$, $T$ and $U$ \cite{Peskin:1990zt} are useful to constraint NP that enters in self-energy corrections to a gauge boson propagator, denoted by $\Pi^{}_{ij}$, which represents the transition $ij$ $(i,j=W,Z,\gamma)$, as  we have \cite{Olive:2016xmw}
\begin{equation}
\hat{\alpha}(M_Z)T=\frac{\Pi^{\rm NP}_{WW}(0)}{M^2_W}-\frac{\Pi^{\rm NP}_{ZZ}(0)}{m^2_Z},
\end{equation}
where $\hat{\alpha}(M_Z)$ is the renormalised Electro-Magnetic (EM) coupling constant at the $M_Z$ scale. Here, we are interested in the $T$ parameter. In this respect, a related quantity known as the $\rho$ parameter is defined as \cite{Olive:2016xmw}
\begin{equation}
\rho-1=\frac{1}{1-\hat{\alpha}(M_Z)T}\simeq\hat{\alpha}(M_Z)T.
\end{equation}
		
In this work we take $\Delta{\rho^{\rm exp}}=\rho-1=0.0006\pm0.0009$,  which is extracted from the data on the $T$ parameter $(0.08\pm0.12)$ \cite{Olive:2016xmw}.  While in the SM $\rho\equiv\rho_0=M_{W^\pm}^2/M_Z^2\cos{\theta_W}=1$ at tree level,  in our scan we obtain $\Delta{\rho}^{\rm SUSY}\in[0.0001,0.0006]$.}} {However, we will focus on the strongest constraint, which is in fact from the decay $\tau\to l \nu_\tau \nu_{l}$, essentially because it carries the same one-loop corrections of the vertex $W^\pm l\nu_l$ within the process $b\to cl\nu_l$.}}  In order to have sizable loop functions, we will enforce on our scans  the condition $m_{\chi^{0}_1}\approx m_{\chi^{-}_1}\lesssim 500~{\rm GeV}$}.
\begin{figure}[!t]
	\centering
           \includegraphics[height=4.75cm,width=7.60cm]{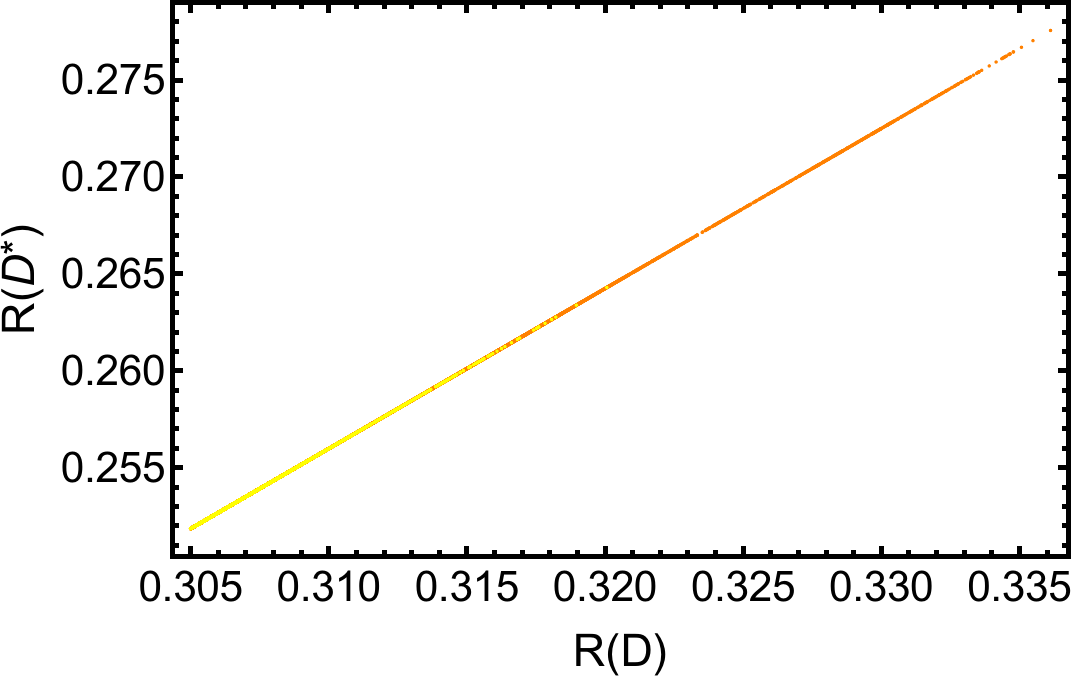}~~~~\includegraphics[height=5cm,width=7.75cm]{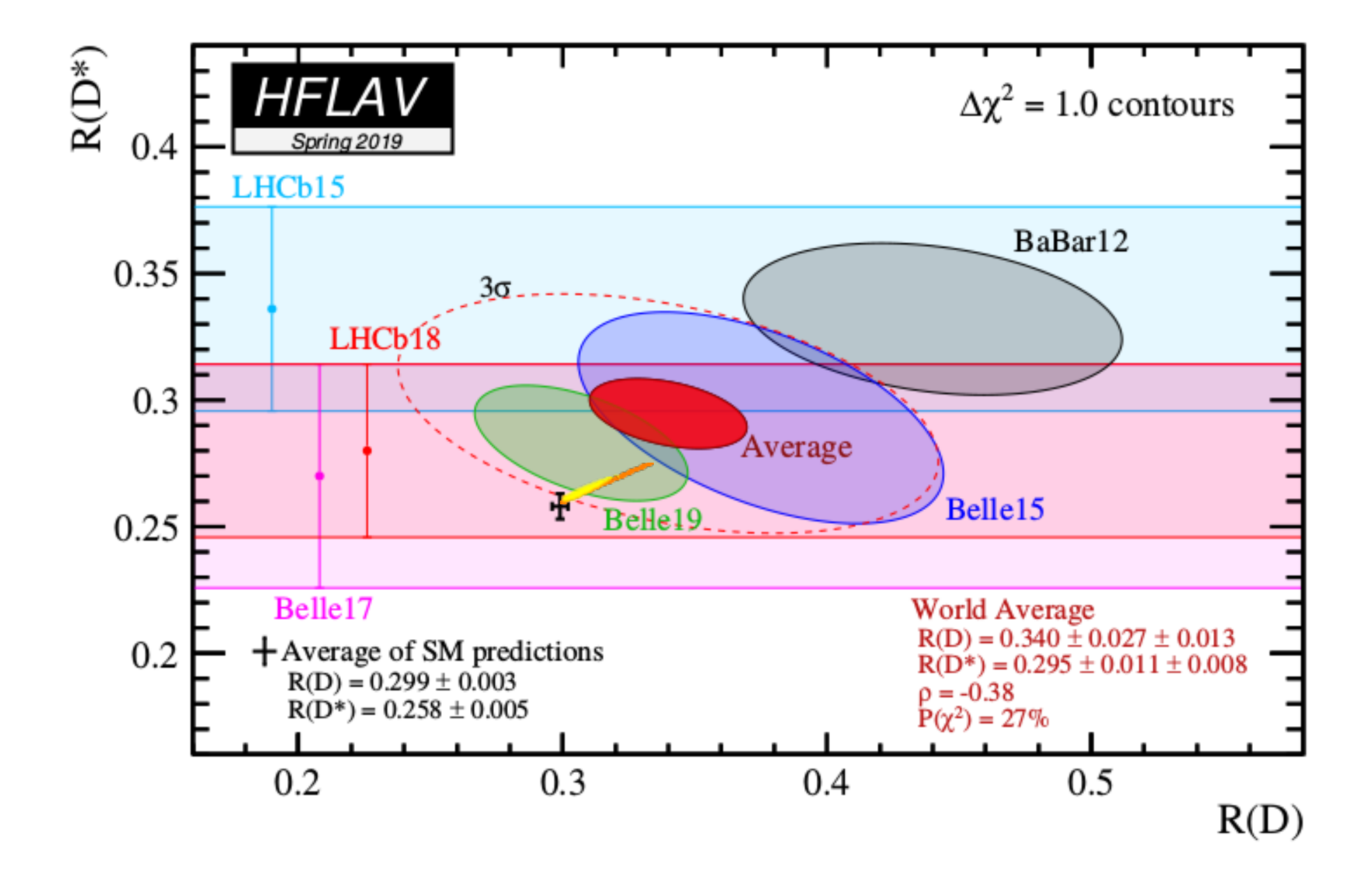}
	\vspace{-0.9cm}
	\caption{(Left) The correlation between ${R}(D)$ and ${R}({D^*})$ after the one-loop SUSY contributions through the lepton penguins where the orange and yellow points show the constrained ones by  $\Gamma(\tau\rightarrow\mu\nu_{\tau}\nu_{\mu})/\Gamma(\tau\rightarrow e\nu_{\tau}\nu_{e})$ at $1\sigma$ and $2\sigma$, respectively. (Right) The comparison between the SUSY contributions shown in the left panel and the new world averages.}
	\label{comb_MSSM_av}
\end{figure}
\begin{figure}[!t]
	\centering
	\includegraphics[height=5.5cm,width=16.2cm]{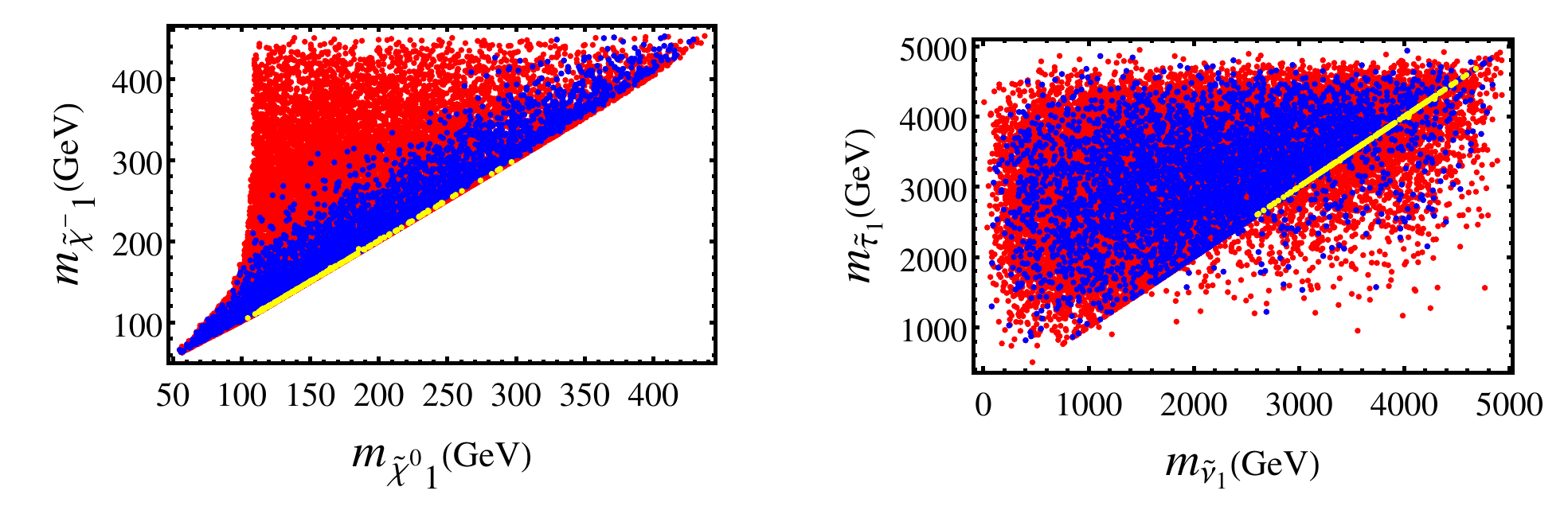}
	\vspace{-0.8cm}
	\caption{The correlation between  chargino and neutralino masses (left) as well as  
sneutrino and stau masses (right). Here,  the blue and red points show the constrained ones by  $\Gamma(\tau\rightarrow\mu\nu_{\tau}\nu_{\mu})/\Gamma(\tau\rightarrow e\nu_{\tau}\nu_{e})$ at $1\sigma$ and $2\sigma$, respectively. The additional yellow points represent the region with ${R}({D})>0.33$.}
	\label{SYSYmasses}
\end{figure}
As mentioned, the enhancement of $C_{VL}^{\tilde{\tau}}$ occurs mostly when the chargino and neutralino masses are light and similar, in addition to large $\tan\beta$ and stau mass. Therefore, in our scan, we focus on benchmark points where the gaugino soft masses {are given  by $M_1$, $M_2$ $\in [110,500]$ GeV and $M_3 =2$ TeV. Also, we choose the $\mu$ parameter $\in[100,500]$ GeV, $m_{A^{0}}^2 \in [0,25\times10^{4}]$ GeV${^2}$, the $A$ terms $\in [-2000,-100]$ GeV, $M_{\tilde{Q}}$, $M_{\tilde{U}}$ and $M_{\tilde{D}}$ are fixed in the TeV range
while the slepton soft mass terms $m_{\tilde{L}}$ and $m_{\tilde{E}} \in [100, 5000]$ GeV. Finally, we take $\tan \beta \in [5, 70]$.
		
In Fig.~\ref{comb_MSSM_av} we present the correlation between ${R}(D)$ and ${R}(D^*)$ at at one-loop due to the SUSY contributions to the lepton penguins alone. As  can be seen from this plot, in presence of MSSM one-loop corrections ,
${R}(D)$ can reach $0.335$ while ${R}(D^*)$ extends to $0.277$ (left panel), which are results rather consistent with the Belle measurements shown by the green ellipse (right frame) and not that far from the BaBar ones. Also, the MSSM one-loop corrections leads  to rather consistent results for ${R}(D)$ (somewhat less so for ${R}(D^*)$) with the averages represented by the red ellipse. {This correlation can be understood from the fact that SUSY one-loop corrections give a significant contribution to $g_{VL}$ only (of order $6\%$) and, hence, according to Eqs. (16)--(17), both $R(D)$ and $R(D^*)$ are affected by the same correction factor $\propto (1+g_{VL})^2$ through a common Wilson coefficient.
	It is also worth noting that the enhancements of $R(D)$ and $R(D^*)$ require a very peculiar region of parameter space of the MSSM, especially in terms of
	$m_{{\tilde\chi}_1^-}$ and $\tan\beta$, wherein, however, all experimental and theoretical constraints sensitive to the latter two quantities are taken into account and included in our scan and numerical analysis.}	
To our knowledge, these enhancements in both  ${R}(D)$ and ${R}(D^*)$ have never been accounted for before in any NP scenario.

It is also very  relevant to extract the typical mass spectra which are responsible for the MSSM configurations yielding
${R}(D)$ and ${R}(D^*)$ values (potentially)
consistent with experimental measurements, as these might be accessible during Run 3 at the LHC. As an indication, this is done in Fig.~\ref{SYSYmasses}  for the case of the chargino and neutralino masses (left frame) as well as  
sneutrino and stau masses (right frame).
The plot shows a predilection of the highest ${R}(D)$ and ${R}(D^*)$ points for MSSM parameter configurations 
with $m_{\tilde\chi_1^\pm}>m_{\tilde\chi_1^0}$
and 
$m_{\tilde\tau_1}>m_{\tilde\nu_1}$ while the absolute mass scale can cover the entire interval from 100 GeV to 400 GeV in the first case and from
200 GeV to 5 TeV in the second case. Further, the points with ${R}(D)>0.33$ prefer both $m_{\tilde\chi_1^\pm}$ and $m_{\tilde\chi_1^0}$ below 300 GeV 
and require a rather large $\tilde\tau_1$ and  $\tilde\nu_1$ masses (say, above 2.5 TeV as well as large $\tan\beta$). This signals that there occurs an interplay between mass suppressions in the loops and enhancements in the couplings.

\section{Conclusion}
We have shown that the MSSM has the potential to explain data by  BaBar and Belle  revealing  rather significant anomalies in $R(D)$ and $R(D^*)$. 
Within this BSM scenario, such excesses can be approached  in presence of  lightest neutralino/chargino mass degeneracy and  
 large $\tilde\tau_1$ and $\tilde\nu_{1}$ masses. Altogether, we found
a more than acceptable agreement with both  Belle (especially) and BaBar (to a lesser extent)  results.


\section*{Acknowledgments}
 DB was supported by the Algerian Ministry of Higher Education and Scientific Research under the PNE Fellowship.
SK acknowledges partial support from the Durham IPPP
Visiting Academics (DIVA) programme.
SM is financed in part through the NExT Institute and the STFC consolidated
Grant No. ST/L000296/1.
The
work of SK and SM was partially supported by the
H2020-MSCA-RISE-2014 grant No. 645722 (NonMinimalHiggs).


\section*{References}

\begin{thebibliography}{9}

\bibitem{Lees:2012xj} 
J.~P.~Lees {\it et al.} [BaBar Collaboration],
Phys.\ Rev.\ Lett.\  {\bf 109}, 101802 (2012).
\bibitem{Lees:2013uzd} 
J.~P.~Lees {\it et al.} [BaBar Collaboration],
Phys.\ Rev.\ D {\bf 88}, no. 7, 072012 (2013).
\bibitem{Huschle:2015rga}
M.~Huschle \textit{et al.} [Belle Collaboration],
Phys.\ Rev.\ D \textbf{92}, no. 7,
072014 (2015).
\bibitem{Sato:2016svk}
Y.~Sato {\it et al.} [Belle Collaboration],
Phys.\ Rev.\ D {\bf 94}, no. 7, 072007 (2016).
\bibitem{Hirose:2016wfn} 
S.~Hirose {\it et al.} [Belle Collaboration],
Phys.\ Rev.\ Lett.\  {\bf 118}, no. 21, 211801 (2017).
\bibitem{Hirose:2017dxl} 
S.~Hirose {\it et al.} [Belle Collaboration],
Phys.\ Rev.\ D {\bf 97}, no. 1, 012004 (2018)
\bibitem{Abdesselam:2019dgh} 
A.~Abdesselam {\it et al.} [Belle Collaboration],

\bibitem{Aaij:2015yra} 
R.~Aaij {\it et al.} [LHCb Collaboration],
Phys.\ Rev.\ Lett.\  {\bf 115}, no. 11, 111803 (2015)
Erratum: [Phys.\ Rev.\ Lett.\  {\bf 115}, no. 15, 159901 (2015)].
\bibitem{Aaij:2017uff} 
R.~Aaij {\it et al.} [LHCb Collaboration],
Phys.\ Rev.\ Lett.\  {\bf 120}, no. 17, 171802 (2018).
\bibitem{Aaij:2017deq} 
R.~Aaij {\it et al.} [LHCb Collaboration],
Phys.\ Rev.\ D {\bf 97}, no. 7, 072013 (2018).

\bibitem{Amhis:2019ckw} 
Y.~S.~Amhis {\it et al.} [HFLAV Collaboration],
arXiv:1909.12524 [hep-ex].
\bibitem{Boubaa:2016mgn} 
  D.~Boubaa, S.~Khalil and S.~Moretti,
  Int.\ J.\ Mod.\ Phys.\ A {\bf 34}, no. 32, 1950209 (2019).
 \bibitem{Bhattacharya:2015ida}
 S.~Bhattacharya, S.~Nandi and S.~K.~Patra,
 Phys.\ Rev.\ D {\bf 93}, 034011 (2016).
\bibitem{Hagiwara:1989cu}
K.~Hagiwara, A.~D.~Martin and M.~F.~Wade,
Nucl.\ Phys.\ B {\bf 327}, 569 (1989).
\bibitem{Hagiwara:1989gza}
K.~Hagiwara, A.~D.~Martin and M.~F.~Wade,
Phys.\ Lett.\ B {\bf 228}, 144 (1989).
\bibitem {Datta:2012qk}
A.~Datta, M.~Duraisamy and D.~Ghosh,
Phys.\ Rev.\ D {\bf 86}, 034027 (2012)

\bibitem{Duraisamy:2013kcw} 
M.~Duraisamy and A.~Datta,
JHEP {\bf 1309}, 059 (2013)

\bibitem{Fajfer:2012vx} 
S.~Fajfer, J.~F.~Kamenik and I.~Nisandzic,
Phys.\ Rev.\ D {\bf 85}, 094025 (2012).

\bibitem{Crivellin:2012ye} 
A.~Crivellin, C.~Greub and A.~Kokulu,
Phys.\ Rev.\ D {\bf 86}, 054014 (2012).

\bibitem{Crivellin:2013wna} 
A.~Crivellin, A.~Kokulu and C.~Greub,
Phys.\ Rev.\ D {\bf 87}, no. 9, 094031 (2013).
\bibitem{Tanaka:2012nw} 
M.~Tanaka and R.~Watanabe,
Phys.\ Rev.\ D {\bf 87}, no. 3, 034028 (2013)

\bibitem{Sakaki:2013bfa} 
Y.~Sakaki, M.~Tanaka, A.~Tayduganov and R.~Watanabe,
Phys.\ Rev.\ D {\bf 88}, no. 9, 094012 (2013)

\bibitem{Sakaki:2014sea} 
Y.~Sakaki, M.~Tanaka, A.~Tayduganov and R.~Watanabe,
Phys.\ Rev.\ D {\bf 91}, no. 11, 114028 (2015).
\bibitem{Buras:2002vd}
A.~J.~Buras, P.~H.~Chankowski, J.~Rosiek and L.~Slawianowska,
Nucl.\ Phys.\ B {\bf 659}, 3 (2003).
\bibitem{Porod:2014xia} 
W.~Porod, F.~Staub and A.~Vicente,
Eur.\ Phys.\ J.\ C {\bf 74}, no. 8, 2992 (2014)

\bibitem{Staub:2013tta}
F.~Staub,
Comput.\ Phys.\ Commun.\  {\bf 185}, 1773 (2014).
\bibitem{Porod:2011nf}
W.~Porod and F.~Staub,
Comput.\ Phys.\ Commun.\  {\bf 183}, 2458 (2012).
\bibitem{Olive:2016xmw}
C.~Patrignani,
Chin.\ Phys.\ C {\bf 40}, no. 10, 100001 (2016).
\bibitem{Aubert:2009qj} 
  B.~Aubert {\it et al.} [BaBar Collaboration],
  Phys.\ Rev.\ Lett.\  {\bf 105}, 051602 (2010).
\bibitem{Peskin:1990zt} 
  M.~E.~Peskin and T.~Takeuchi,
  Phys.\ Rev.\ Lett.\  {\bf 65}, 964 (1990).
\end{thebibliography}
\end{document}